# Fast Feature Reduction in Intrusion Detection Datasets

Shafigh Parsazad*, Ehsan Saboori**, Amin Allahyar*

*Department Of Computer Engineering, Ferdowsi University of Mashhad, Mashhad, Iran.
**K.N Toosi University of Technology, Tehran, Iran.
Shafigh.Parsazad@stu-mail.um.ac.ir, ehsan.saboori@ieee.org, amin.allahyar@stu-mail.um.ac.ir

*Abstract*— In the most intrusion detection systems (IDS), a system tries to learn characteristics of different type of attacks by analyzing packets that sent or received in network. These packets have a lot of features. But not all of them is required to be analyzed to detect that specific type of attack. Detection speed and computational cost is another vital matter here, because in these types of problems, datasets are very huge regularly. In this paper we tried to propose a very simple and fast feature selection method to eliminate features with no helpful information on them. Result faster learning in process of redundant feature omission. We compared our proposed method with three most successful similarity based feature selection algorithm including *Correlation Coefficient, Least Square Regression Error* and *Maximal Information Compression Index.* After that we used recommended features by each of these algorithms in two popular classifiers including: Bayes and KNN classifier to measure the quality of the recommendations. Experimental result shows that although the proposed method can't outperform evaluated algorithms with high differences in accuracy, but in computational cost it has huge superiority over them.

Keywords; intrusion detection; fast feature selection; computational cost; KDD 99.

## I. INTRODUCTION

As the time passes the internet breaches more and more in human life and merge itself in all aspect it. Despite the benefit of internet in making life easier as well as doing varies of works faster than ever, it made the whole new ways to do illicit behavior. These behaviors vary from infiltration to anything that is connected to it illegally to making a resource so busy that it can't service to legitimate users. There are two type of defense mechanism against these attacks: static and dynamics. Security update and firewalls is some example of static defense mechanism. A very popular dynamic defense is *Intrusion Detection Systems* (IDS). In definition, "intrusion means violation of security policy"[1]. With this definition intrusion detection express the concept of finding a way to detect any illegal behavior that is performing in the network or more accurately in transmitting packets. Process of finding these illegal activities assume that normal packets are different from packets that doing intrusive behavior. Intrusion detection is not an alternative to prevention-based techniques (for ex. authentication and access control), instead, it is used to complete the security measures and recognition of actions that tries to bypass the security sections and monitoring of the system. There for Intrusion detection stand in the next level of security component of the system[1]. In short IDS are used as a dynamic methodology to supplement static defense and make it more powerful and accurate. An example of these systems is represented in figure.1. Some specific examples of intrusions that concern system administrators include[2]:

- "Unauthorized modifications of system files so as to facilitate illegal access to either system or user information.

- Unauthorized access or modification of user files or information.

- Unauthorized modifications of tables or other system information in network components (e.g. modifications of router tables in an internet to deny use of the network).

- Unauthorized use of computing resources (perhaps through the creation of unauthorized accounts or perhaps through the unauthorized use of existing accounts)".

Some of the important features an intrusion detection system should possess include[1]:

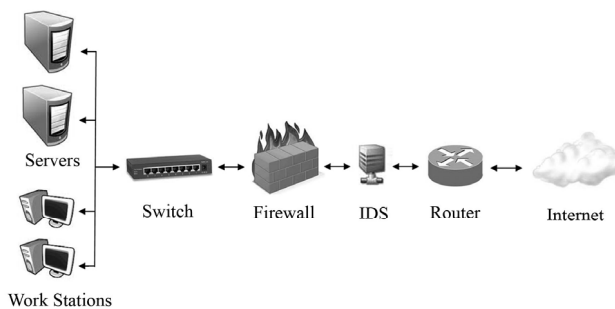

Figure1. Location of an intrusion detection system (IDS) in a network and utilization of it as second line of defense.

- "Be fault tolerant and run continually with minimal human supervision. The IDS must be able to recover from system crashes, either accidental or caused by malicious activity.

- Possess the ability to resist subversion so that an attacker cannot disable or modify the IDS easily. Furthermore, the IDS must be able to detect any modifications forced on the IDS by an attacker.

- Impose minimal overhead on the system to avoid interfering with the normal operation of the system.

- Be configurable so as to accurately implement the security policies of the systems that are being monitored. The IDS must be adaptable to changes in system and user behavior over time.

- Be easy to deploy: this can be achieved through portability to different architectures and operating systems, through simple installation mechanisms, and by being easy to use by the operator.

- Be general enough to detect different types of attacks and must not recognize any legitimate activity as an attack (false positives). At the same time, the IDS must not fail to recognize any real attacks (false negatives)".

There are varieties of attacks that IDS tries to detect. Some of these can be detected by scanning the packets to find signature of specific attacks. Other types of attacks are very much like normal packet pattern with slight difference in packet content. So there is two type of intrusion detection algorithm. They are called Misuse Detection and Anomaly Detection respectively.

*A. Anomaly Detection*

Anomaly detection is based on assumption that an attack will always similar to normal pattern with a slight difference. With this assumption a classifier first trained with normal and abnormal samples. It should predict whatever a future coming data should be placed in normal class or abnormal class. The training process is performed with finding characteristics of normal class and definition of rules and heuristics depend on observations. A very popular method for doing this is *Support Vector Data Description* (SVDD)[3]. Experimental result shows a very high detection rate and accuracy using SVDD [4-8]. Another approach is Neural Networks[9-10]. Hybrid methods also used to utilize the positive aspects of these two approaches while reducing negative aspects of them as much as possible. SVDD maximizes the margin but it can't handle missing values correctly. Neural networks have O(1) in prediction and ability of handling missing values but it's not obvious what is the optimal weight or how many level is required to achieve the best result. An example of hybrid method is proposed in [11].

*B. Misuse detection*

Misuse detection is based on searching for signature of known attacks in packets that are transmitted on the network. In the other word, misuse detection tries to find intruder who is trying to breach in the system using some known attack type. Although in real world, system administrators should be familiar with all known vulnerability of the system and fix them. In this methodology a term *intrusion scenario* is widely used. For any attack to be performed, there are a certain steps that an intruder will do to make that attack, these steps is called intrusion scenario. Misuse intrusion detection persistently scans recent activities and looking for known intrusion scenario to ensure no one is trying to attack the system. For intrusion detector to be able to detect known attack types, every attack should be represented or modeled in specific ways that the system can recognize it. This is the point where different methodologies for misuse detection are varying[1].

II. PREPROCESSING AND DATA PROVISION

*A. Preparing Raw Network Data*

To performing learning process packets from network are collected and fed to the learning algorithm. Regularly these networks transmits huge amount of data every second. So that, datasets for intrusion detection are very populated and extensive amount of data should be analyzed. Also packets transmitting on the network have a lot of overhead on them that is not related to the content of the packet. Because of this a preprocessing is performed for raw network data. In the other word raw network packets is converted to structures that have closer level to meaning of that connection, for example connection records. Any of these structures have set of features describing them. Deciding which of these features can represent more knowledge is a critical decision and one should have very wide and accurate knowledge about the domain to do that. Because of this, there are not a lot of dataset for intrusion detection. As far as the authors know there are only two dataset for intrusion detection: KDD 99[12] and IDEval[13], while the former is only a refinement version of the later one.

*B. KDD 99*

The KDD 99 is an intrusion detection dataset that is build upon about four gigabytes of compressed binary TCP dump data from seven weeks of network traffic. This was processed into about five million connection records. Similarly, the two weeks of test data yielded around two million connection records. The network was a simulation of military network which had three servers with rule of victim computers that exposure a lot of attacks and normal network traffic. All attacks were consisted on four main categories:

- Denial of Service (dos): Intruder tries to consume server resources as much as possible, so that normal users can't get resources they need.

- Remote to Local (r2l): Intruder has no legitimate access to victim machine but tries to gain access.

- User to Root (u2r): Intruder has limited privilege access to victim machine but tries to get root privilege.

- Probe: Intruder tries to gain some information about victim machine.

The original raw dump network traffic were preprocessed for International Knowledge Discovery and Data Mining Tools

Competition[14]. For this, all the raw data is converted into connection. A connection means "a sequence of TCP packets starting and ending at some well defined times, between which data flows from a source IP address to a target IP address under some well defined protocol"[15]. This conversion is performed by Paxson algorithm[16]. Result of this algorithm has 41 features. Each of features is member of these four categories[17]:

- Basic Features: These features are captured from packet headers only and without analyzing payload. Features 1 to 6 are in this category.

- Content Features: In this category original tcp packets analyzed with assistance of domain knowledge. An example of this category is number of "hot" indicators.

- Time-based Traffic Features: for capturing these types of features a window of 2 second interval is defined. In this interval, some properties of packets is measured. For example number of connections to the same service as the current connection in the past two seconds.

- Host-based Traffic Features: In this category instead of a time based window, a number of connections are used for building the window. This category is designed so that attacks longer than 2 second can be detected.

The KDD 99 is distributed in three versions which are described in table.1. "10% KDD" has 22 attack types and is a synopsis version of "Whole KDD" dataset. These datasets has more attack samples than normal samples. Because of specific attribute of DoS attacks, it has more frequency in dataset than other types. In the "Corrected KDD" dataset distribution of data is changed also it has more 14 additional attacks. The list of subcategories of "10% KDD" is represented in table.2. This version of dataset is used for wide variety of studies and more importantly it is used in the contest. So for evaluation result, we used this version. This version of dataset is used for wide variety of studies and more importantly it is used in the contest. So for evaluation result, we used this version. Feature values for each sample in this dataset have a very unbalanced range in compare of other features. For example feature #1 has range between 0 and 42448 and feature #10 has range between 0 and 28. Because of this performing normalization on dataset before training phase seems reasonable although this process can be done in preprocessing phase too.

TABLE I. SAMPLE DISTRIBUTION IN EACH TYPE OF KDD 99 DATASET VERSION.

| Dataset | Packet Category | | | | |
|---|---|---|---|---|---|
| | *DoS* | *Probe* | *U2R* | *R2L* | *Normal* |
| 10% KDD | 391458 | 4107 | 52 | 1126 | 97277 |
| Corrected KDD | 229853 | 4166 | 70 | 16347 | 60593 |
| Whole KDD | 3883370 | 41102 | 52 | 1126 | 972780 |

TABLE II. TABLE 1. LIST OF SUBCATEGORIES IN "10% KDD".

| Attack | Number of Samples | Category |
|---|---|---|
| smurf | 280790 | dos |
| neptune | 107201 | dos |
| back | 2203 | dos |
| teardrop | 979 | dos |
| pod | 264 | dos |
| land | 21 | dos |
| normal | 97277 | normal |
| satan | 1589 | probe |
| ipsweep | 1247 | probe |
| portsweep | 1040 | probe |
| nmap | 231 | probe |
| warezclient | 1020 | r2l |
| guess_passwd | 53 | r2l |
| warezmaster | 20 | r2l |
| imap | 12 | r2l |
| ftp_write | 8 | r2l |
| multihop | 7 | r2l |
| phf | 4 | r2l |
| spy | 2 | r2l |
| buffer_overflow | 30 | u2r |
| rootkit | 10 | u2r |
| loadmodule | 9 | u2r |
| perl | 3 | u2r |

III. SIMILARITY MEASURES

In this section we introduce some of most successful similarity measures. These successful similarity measures includes: *Correlation Coefficient[18], Least Square Regression Error[19-22]* and *Maximal Information Compression Index[23]*.

A. *Correlation Coefficient[18]:*

One of the most popular and known measures of similarity between two random variable is correlation coefficient. It also called cross-correlation coefficient. We can use (1) to measure how much random variable $x$ is similar to random variable $y$.

$$Relation(x,y) = \frac{cov(x,y)}{\sqrt{var(x) \times var(y)}} \quad (1)$$

Where $var(x)$ correspond to variance of variable $x$ and $cov(x,y)$ correspond to covariance of random variable $x$ and $y$. Covariance of two variables shows strength of relation between those variables. If $x$ grow when $y$ rises with exact linear relation, covariance of these two variable will be $+1$ and if $x$ shrink when $y$ rises covariance will be $-1$. If two variables are unrelated at all covariance will be 0. We can use this as a similarity measure. This measure have following properties[23]:

1. $0 \leq 1 - |Relation(x,y)| \leq 1$.

2. $1 - |Relation(x,y)| = 0$ If and only if $x$ and $y$ are linearly related.

3. This measure is symmetric. Meaning that always $Relation(x, y) = Relation(y, x)$.
4. Assume $u = \frac{x-a}{c}$ and $v = \frac{y-b}{d}$. For any arbitrary value for $a, b, c, d$ we have:

$$|Relation(x, y)| = |Relation(u, v)|$$

5. $Relation(x, y)$ Is *sensitive to rotation* of the scatter diagram in $(x, y)$ plane.

As Mitra[23] discussed correlation coefficient have a lot of convenient specifications but because of having properties of 4 and 5 mentioned above it is invariant to scale. Because of this, it's possible that two variables have same similarity measure although they have very different variances. It is not preferable because variance always gives high information about content.

*B. Least Square Regression Error:*

Another similarity measure that is used regularly is least square regression error or *residual variance*. It is the error of predicating $y$ from $y = bx + a$. Where $a$ and $b$ are the regression coefficients which can be calculated by minimizing $e(x, y)^2$ in (2):

$$e(x, y)^2 = \frac{1}{n}\sum e(x, y)_i^2 \quad (2)$$

Where $e(x, y)_i$ can be calculated with (3):

$$e(x, y)^2 = y_i - a - bx_i \quad (3)$$

So we have (4), (5):

$$a = \overline{y} \quad (4)$$

$$b = \frac{cov(x, y)}{var(x)} \quad (5)$$

Having $a$ and $b$ we can calculate mean square error $e(x, y)$ by (6):

$$e(x, y) = var(y) \times (1 - Relation(x, y)^2) \quad (6)$$

With this new measure if $x$ and $y$ have linear relation $e(x, y)$ will be 0 and if $x$ and $y$ have no relation at all $e(x, y)$ will equal to $var(x)$. Mitra also argues about properties of this measure[23]:

1. $0 \leq e(x, y) \leq var(y)$.
2. $e(x, y)$ Is equal to 0 if and only if $x$ and $y$ have perfect linear relation.
3. This measure is not symmetric.
4. Assume $u = \frac{x}{c}$ and $v = \frac{y}{d}$. For any arbitrary value for $c, d$ we have $e(x, y) = d^2 e(u, v)$. With this property we can conclude that $e$ is *sensitive to scaling* of the variables. It's obvious that $e$ is not *sensitive to translation*.
5. $e$ Is *sensitive to rotation* of scatter diagram in $x - y$ plane.

*C. Maximal Information Compresseion Index[23]:*

This measure fixed some defects of two previous described measures. We refer to it by MICI and use $\lambda_2$ for simplicity. MICI can be calculated using (7):

$$\lambda_2(x, y) = \min(eigv(\Sigma)) \quad (7)$$

Where $\Sigma$ correspond to covariance of $x$ and $y$ and $eigv$ is a vector of eigen values of that covariance.

The $\lambda_2$ will equal to 0 when the feature have linear relation and as much as the relation fade away, the value of $\lambda_2$ increase. As the formula represent $\lambda_2$ is only an eigenvalue for a direction where data have the most elongation. In the other word it is the main idea behind *Principle Component Analysis* (PCA)[20, 24]. PCA is based on the fact that if a data be projected along its principal component direction, it yields maximum information compaction. If we reconstruct the data, the amount of lost information is related to eigenvectors that is not considered in PCA calculation. In PCA we always use $k$ eigenvectors that have maximum corresponding eigenvalues. So the amount of lost information will be minimal. The $\lambda_2$ have following properties[23]:

1. $0 \leq \lambda_2(x, y) \leq 0.5(var(x) + var(y))$.
2. $\lambda_2(x, y)$ Will equal to 0 if and only if $x$ and $y$ is linearly related.
3. $\lambda_2(x, y) = \lambda_2(y, x)$, so it's a symmetric measure.
4. Assume $u = \frac{x}{c}$ and $v = \frac{y}{d}$. For any arbitrary value for $c, d$ we have $\lambda_2(x, y) \neq \lambda_2(u, v)$ so it is *sensitive to scaling* and is not related to *translation*.
5. $\lambda_2(x, y)$ Is not related to rotation.

To detect which feature should be removed from dataset using discussed similarity measures, a simple general algorithm based on *K-Nearest Neighbor* (KNN) algorithm[25] is used. It has two phases. In the first phase, all of features partitioned using the similarity measure as a distance in KNN algorithm. In the second phase a feature that describes its cluster more accurately is selected as a candidate of that group.

## IV. PROPOSED METHOD

As discussed earlier, variance of a random variable has very high information inside. Intuitively, we used this measure to find features with low quality to be eliminated from dataset while preserving others. In the other word we used variance to discover if any feature has no class dependent information included. Features with low information should be removed. Also because of nature of intrusion detection datasets, they are very populated by samples and even a linear order for computational cost is very slow. So we tried to make our approach to be simple as possible while preserving its accuracy.

If we look precisely at a simple two class dataset with separate classes, we can clearly see that if there is any feature in dataset that have separate range of values in each class, we can classify all of samples in the dataset with 100% accuracy with a linear classifier. An example of such a situation is shown in figure.2. Now let's assume a more general but still simple dataset shown in figure.3. In this dataset it is clearly obvious that if we remove any features from dataset, the accuracy of classification will not be 100% although the classes are separated with high distance. In real world dataset usually we have a very mixed dataset but we can still use this fact to detect if any feature can have more roles on separation of classes than other features. We developed this fact and built our algorithm upon it.

Let $X = \{x_i\}_{i=1}^{N}$ be our intrusion detection dataset, containing $N$ network connection samples containing all attack and normal connections. Each $x_i$ is a vector with $D$ dimension and $x_i \in \mathbb{R}^D$. Let $L = \{l_i\}_{i=1}^{N}$ be our class labels for each sample. Also we have $l_i \in k$ where $k = \{1, ..., K\}$. With this definition we have $K$ class in our dataset and $K = 5$ because we used "KDD 10%" as our dataset. $x_{id}$ Represent the value of feature $d = \{1, ..., D\}$ in connection sample $i$. For each class first we add all data point $x_i$ which are in same class, feature by feature (8):

$$S_{cd} = \left\{ for\ all\ x_i\ where\ \ l_i = c \Big| \frac{\sum x_{id}}{n_c} \right\} \quad (8)$$

Where $n_c$ is number of samples in class $c$ and $S_{cd}$ is mean of connection sample $x_i$ along feature $d$ whose class is $c$ and $c \in k$. We put all $S_{cd}$ in a vector and call it $FS$(9):

$$S_d = [S_{1d}\ \ S_{2d}\ ...\ S_{Kd}] \quad (9)$$

After that we need to calculate how much $S_d$ is scattered along each feature. In the other word we need to measure confusion of $S_d$ for each feature. For this reason we used variance of $S_d$(10):

$$V_d = Var(S_d) = E[(S_d - \mu_d)^2] \quad (10)$$

Where $\mu_d$ is mean of all values in $S_d$ and $V_d$ is variance of $S_d$. Items in $V = \{V_d\}_{d=1}^{D}$ with high value represent degree of turbulence in feature $d$ among all classes. High turbulence represents power of corresponding feature in class separation. Now we can sort $V$ with ascending order and select $t$ feature index for classification where $t$ is number of desired feature that user defined to be selected.

## V. EVALUATION

In this section we used KDD 99 dataset to evaluate these three similarity measure and compare them with our proposed method. To survey the effectiveness of these feature selection algorithm in classification, we used two popular classification algorithms including: Bayesian Network Classifier[26-27] and K-nearest neighbor classifier[25] and measured their accuracy.

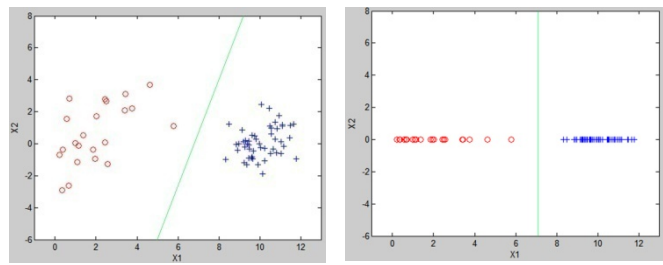

Figure 2. Demonstration of the fact that we can still classify data with 100% accuracy with only one feature when we have separate range of data on that feature corresponds to its class.

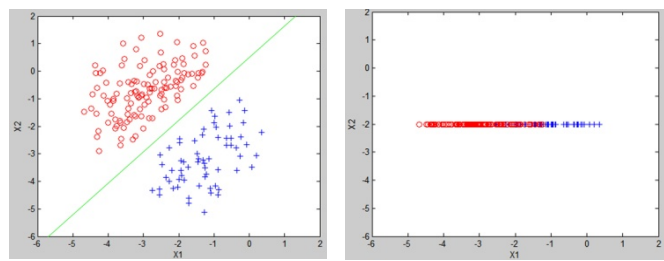

Figure 3. Demonstration of the fact that we can't classify the data with 100% accuracy when there is no feature with separate range corresponds to its class.

Evaluation is done with 10-fold algorithm to ensure the result is as authentic as possible. So the goal for these feature selection algorithm is to keep the class separability of the dataset as much as possible while reducing its size to lower dimension. Also selection algorithm should be very fast because in real world problem we encounter very large datasets and slow algorithm is not applicable.

First of all we classified the original "10% KDD" dataset with KNN and Bayes classifiers, result of this process is represented in first main column of table.3 and table.4. After that, dataset is fed to three discussed methodology also to our proposed method. Then the original dataset is reduced to features recommended by each of these similarity measure based algorithms and FFR (Fast Feature Reduction). At last the reduced dataset is again fed to KNN and Bayes classifier and result of this classification measured.

In this research speed of feature selection was our main goal. Because of this consumed time of each algorithm with different reject threshold measured and shown in table.3. The evaluated result for each algorithm is represented in its respective column of table.4 and table.5. As evaluation result shows, although FFR cannot defeat other methodologies in accuracy of classification and accuracy didn't changed very much, but in speed FFR outperformed all other feature selection method with great differences.

TABLE III. FEATURE SELECTION CALCULATION TIME IN SECOND.

| | CC (Correlation Coefficient) | | | LSRE (Least Square Regression Error) | | | MICI (Maximal Information Compression Index) | | | FFS (Fast Feature Selection) | | |
|---|---|---|---|---|---|---|---|---|---|---|---|---|
| Features size | 10 | 20 | 30 | 10 | 20 | 30 | 10 | 20 | 30 | 10 | 20 | 30 |
| Speed | 0.300 | 0.308 | 0.318 | 1.464 | 1.473 | 1.480 | 1.766 | 1.778 | 1.797 | **0.006** | **0.008** | **0.010** |

TABLE IV. EVALUATION ACCURACY RESULT OF KNN CLASSIFIER IN PRESENT.

| | | All | CC | | | LSRE | | | MICI | | | FFS | | |
|---|---|---|---|---|---|---|---|---|---|---|---|---|---|---|
| Features size | | *41* | *10* | *20* | *30* | *10* | *20* | *30* | *10* | *20* | *30* | *10* | *20* | *30* |
| Attack Type | **Normal** | 98.20 | 32.93 | 96.33 | **98.65** | 71.04 | 98.00 | 98.00 | 98.10 | 97.85 | 97.85 | **98.20** | **98.25** | 98.30 |
| | **DoS** | 98.47 | 29.09 | 97.51 | **98.34** | 63.67 | 93.44 | **98.34** | 85.20 | 96.54 | 98.03 | **98.28** | **98.28** | 98.09 |
| | **Prob** | 98.62 | 55.07 | 96.41 | 98.38 | 75.14 | **98.98** | **98.50** | 98.92 | 98.23 | **98.65** | 98.50 | 98.42 | 98.42 |
| | **R2L** | 97.88 | 16.33 | 90.91 | 97.03 | 87.01 | 97.62 | 97.43 | **98.04** | 98.05 | **97.80** | 97.61 | **97.79** | 97.61 |
| | **U2R** | 82.00 | **83.33** | 65.79 | 76.47 | 56.00 | **82.00** | 82.61 | 77.55 | 88.64 | **90.70** | 82.00 | 79.59 | 79.25 |
| Over all | | 98.24 | 32.54 | 95.55 | 98.07 | 71.35 | 97.09 | 98.06 | 94.92 | 97.66 | **98.11** | **98.14** | 98.11 | 98.04 |

TABLE V. EVALUATION ACCURACY RESULT OF BAYES CLASSIFIER IN PRESENT.

| | | All | CC | | | LSRE | | | MICI | | | FFS | | |
|---|---|---|---|---|---|---|---|---|---|---|---|---|---|---|
| Features size | | *41* | *10* | *20* | *30* | *10* | *20* | *30* | *10* | *20* | *30* | *10* | *20* | *30* |
| Attack Type | **Normal** | 56.26 | 34.30 | 49.67 | **70.30** | 51.60 | 50.72 | 58.79 | **82.94** | 57.85 | 57.26 | 51.52 | 55.87 | 53.94 |
| | **DoS** | 94.50 | NaN | **98.71** | 93.89 | 32.98 | 85.08 | **95.15** | 30.15 | 60.67 | 93.13 | **61.07** | 90.97 | 93.24 |
| | **Prob** | 94.92 | 68.27 | **92.85** | 93.73 | 81.88 | 92.17 | 93.71 | 65.27 | 78.96 | **93.81** | 75.07 | 89.16 | 92.74 |
| | **R2L** | 91.04 | 47.14 | 60.17 | 63.86 | **99.65** | 90.43 | 91.12 | 77.21 | 93.46 | 87.72 | 92.22 | **93.86** | 95.10 |
| | **U2R** | 32.73 | **33.33** | 8.33 | 23.33 | 2.90 | 3.90 | 21.15 | 3.83 | 3.93 | 19.49 | 8.06 | **15.03** | 54.55 |
| Over all | | 76.09 | 44.17 | 69.87 | **79.29** | 35.42 | 55.15 | 77.22 | 31.56 | 55.75 | 75.73 | **54.87** | **71.94** | 73.91 |